\documentclass[journal]{IEEEtran}

\usepackage{graphicx,mathtools}
\usepackage{amsfonts}
\usepackage{tabularx}
\usepackage[utf8]{inputenc}
\usepackage{amsmath}
\usepackage{amssymb}
\usepackage{ifoddpage}
\usepackage{caption}
\usepackage{subcaption}
\usepackage{lipsum}
\usepackage{graphicx}
\usepackage{xcolor}
\usepackage{cite}
\usepackage{mathtools,leftidx} 
\usepackage{booktabs}
\usepackage{multirow}
\usepackage{makecell}
\usepackage{algpseudocode}
\usepackage[linesnumbered,ruled,vlined]{algorithm2e}

\usepackage{dblfloatfix}    
\usepackage{kotex}
\usepackage{verbatim}

\usepackage{diagbox}
\newcolumntype{C}{>{\centering\arraybackslash}X}
\usepackage[export]{adjustbox}
\usepackage{ifoddpage}
\makeatletter
\def\hlinewd#1{%
\noalign{\ifnum0=`}\fi\hrule \@height #1 %
\futurelet\reserved@a\@xhline}
\makeatother
\ifCLASSINFOpdf
\else
\fi

\usepackage{filecontents}

\usepackage{url}

\begin{document}

\title{Multidimensional Swarm Flight Approach For Chasing Unauthorized UAVs Leveraging Asynchronous Deep Learning}

%
%
%
\author{Tae-Won~Ban, \IEEEmembership{Member, IEEE}, Kyu-Min~Kang, \IEEEmembership{Non-Member}, and Bang Chul Jung, \IEEEmembership{ Senior Member}
\thanks{T.-W. Ban is with Department of Artificial Intelligence and Information Engineering, Gyeongsang National University, Jinju 52828, South Korea (e-mail: twban35@gnu.ac.kr).}
\thanks{K.-M. Kang is with Radio Research Division, Electronics and Telecommunications Research Institute, Daejeon, 34129 South Korea (e-mail:kmkang@etri.re.kr).}

}
%
%
\markboth{Submitted to IEEE Transactions on Vehicular Technology}
{Shell \MakeLowercase{\textit{et al.}}: Bare Demo of IEEEtran.cls for IEEE Journals}
%
%
\maketitle
\begin{abstract}
This paper introduces a novel unmanned aerial vehicles (UAV) chasing system designed to track and chase unauthorized UAVs, significantly enhancing their neutralization effectiveness. The system utilizes a multidimensional swarm flight strategy, employing deep reinforcement learning (DRL) to dynamically adapt the tracking unit's movements based on the received signal strength indicators (RSSIs) emitted by unauthorized UAVs. Asynchronous learning techniques involving multiple agents are implemented to expedite the system’s learning process. A key feature of our approach is the coordinated use of a swarm of UAVs, which circumvents the considerable size burden associated with mounting multiple antennas on a single UAV. We further refine the asynchronous DRL framework by integrating advanced channel modeling techniques, such as spatial correlation and Doppler shift, to augment the robustness and adaptability of the system. Performance evaluations confirm the system's efficacy under varying channel conditions and operational scenarios. Key contributions include the integration of tracking and chasing functionalities into a unified system, the employment of realistic channel models to enhance system adaptability, and a comprehensive analysis of the relationship between channel sampling frequency and chasing performance. This research advances the field of UAV regulation and control, offering a scalable and effective solution to the escalating security challenges posed by unauthorized UAVs.
\end{abstract}

\IEEEpeerreviewmaketitle

\section{Introduction}

\subsection{Background and Motivation}
\IEEEPARstart{O}{riginally} developed primarily for military applications, unmanned aerial vehicles (UAVs) have expanded into a wide array of civilian applications, including agriculture, environmental monitoring, disaster response, and film production \cite{droneapp2021}. Supported by significant advancements in UAV technology \cite{Martinez2018, Cisar2020}, these developments include miniaturization, enabling the creation of smaller, more versatile UAVs that are easily integrated into various daily tasks.

Additionally, UAVs increasingly incorporate artificial intelligence (AI) to enhance their autonomy, significantly improving operational efficiency and reducing the need for human oversight. Further advancements in battery technology and propulsion systems have extended UAV endurance and range, supporting missions that require longer flight times and greater distances. These technological improvements have spurred rapid growth in the UAV market, projected to increase from 26.2 billion USD in 2022 to 38.3 billion USD in 2027, at a compound annual growth rate (CAGR) of 7.9\% \cite{MarketsandMarkets}.

However, the increasing deployment of UAVs across various sectors raises concerns about security, privacy, and airspace regulation. Unauthorized UAVs pose risks ranging from accidental airspace intrusions to deliberate threats such as espionage and terrorism. To mitigate these risks, robust anti-UAV technologies are needed to ensure airspace safety and compliance \cite{Xu2020}. The development of anti-UAV systems has mainly focused on identifying \cite{Tedeschi2024, FAA2023}, detecting \cite{Sadovskis2022, Zhang2023, You2023, Lee2023}, and tracking \cite{He2023, An2024, Jin2024, KianiGaloogahi2017, Danelljan2016, Li2023a, Li2022, Li2023} unauthorized UAVs. However, most conventional approaches are limited to localizing unauthorized UAVs rather than actively chasing them to support the efficient neutralization.

\subsection{Problem Statement and Objective}

Despite significant progress in UAV tracking technologies, existing methods generally focus on monitoring unauthorized UAVs without providing real-time chasing and interception capabilities. Traditional approaches often suffer from limitations in dynamic environments where UAVs exhibit unpredictable motion and evade detection. Especially, traditional feedback-based control methods such as PID control or optimal control techniques typically rely on predictable system dynamics or cooperative control laws, which assume that the target follows a known motion model or provides feedback. Thus, they are not suitable for unauthorized UAV chasing systems. This paper addresses these limitations by proposing a novel multidimensional swarm flight approach for actively chasing unauthorized UAVs, leveraging deep reinforcement learning (DRL). Unlike conventional tracking methods, the proposed approach enables UAVs to pursue unauthorized UAVs in real-time, significantly improving security measures. The primary objectives of this work are as follows:

\begin{itemize}
    \item Achieve real-time pursuit and interception of unauthorized UAVs, beyond just tracking their locations.
    \item Optimize control strategies for coordinated UAV swarms, ensuring rapid response and adaptability.
    \item Reduce tracking errors and computational complexity by utilizing asynchronous learning methods.
    \item Improve resilience to environmental variations by incorporating spatial correlation and Doppler effects into the control framework.
\end{itemize}
These objectives collectively enhance the effectiveness of UAV interception systems, minimizing security threats while ensuring efficient airspace management.

\subsection{Previous Works}

The Federal Aviation Administration (FAA) recently introduced a rule known as remote identification, aimed at increasing accountability in UAV operations by requiring UAVs to broadcast sensitive data, such as their identity \cite{Tedeschi2024, FAA2023}. However, this measure can be circumvented by rogue or unauthorized UAVs.

Detecting unauthorized UAVs is crucial for anti-UAV systems \cite{Sadovskis2022}. Primary detection techniques \cite{Zhang2023, You2023, Lee2023} include:
\begin{itemize}
  \item Radar systems that analyze radio waves reflected by UAVs.
  \item Acoustic sensors that identify UAVs based on rotor noise signatures.
  \item Visual or image detection methods using cameras and infrared sensors, often enhanced by deep neural networks.
\end{itemize}

Tracking involves continuously monitoring the trajectory of detected UAVs using various technologies such as radar, acoustic sensors, visual analysis, GPS, and RF signal tracking, to maintain precise control over airspace.

He et al. proposed a method combining coherent and non-coherent processing to enhance micro-UAV tracking under poor signal-to-noise ratios \cite{He2023}. An et al. developed a novel direction-of-arrival (DOA) estimator that improves tracking of unresolved scattering centers \cite{An2024}. Jin et al. examined the effects of location errors on radar-based UAV detection and introduced an adaptive beam control scheme to counteract jamming inaccuracies \cite{Jin2024}.

Significant advancements in UAV tracking include:
\begin{itemize}
  \item Background-aware correlation filters by Kiani Galoogahi et al. that adjust tracking models according to environmental changes \cite{KianiGaloogahi2017}.
  \item Continuous convolution operators by Danelljan et al. for fine-grained UAV tracking \cite{Danelljan2016}.
  \item A learning response interference suppression (RIS) correlation filter by Li et al. to enhance consistency and minimize distractions during UAV tracking \cite{Li2023a}.
\end{itemize}
Li et al. also proposed a computationally efficient tracking algorithm validated using fixed-wing UAV video data \cite{Li2022}, and a general architecture for UAV-to-UAV detection and tracking within a space-air-ground integrated network (SAGIN) was introduced to provide extensive, cost-effective tracking \cite{Li2023}.

Neutralization strategies include net guns, RF jammers, and laser systems to incapacitate or disable rogue UAVs, ensuring the security of airspace \cite{Park2021, Jurn2021}.

\subsection{Contributions of This Work}

To overcome the limitations of conventional tracking methods, this paper presents a novel UAV swarm-based chasing system that bridges the gap between tracking and active pursuit. Unlike traditional approaches that merely localize unauthorized UAVs, our method enables real-time chasing through dynamic and adaptive control strategies. The key contributions of this work are as follows:
\begin{itemize}
\item Development of an integrated tracking and chasing system using asynchronous DRL to actively pursue and intercept unauthorized UAVs, rather than just tracking their locations. This represents a significant shift from passive surveillance to active interception.
\item Introduction of a swarm flight strategy leveraging multiple UAVs to mitigate signal similarity issues associated with mounting multiple antennas on a single UAV. By distributing antennas across multiple UAVs, the system enhances tracking precision and chasing effectiveness.
\item Implementation of advanced channel modeling techniques incorporating spatial correlation and Doppler shift, enhancing system robustness. These modeling techniques enable the UAV swarm to adapt to varying environmental conditions and maintain effective pursuit of rogue UAVs.
\item Comprehensive analysis of the relationship between channel sampling frequency and chasing performance, ensuring optimized control strategies for real-time UAV pursuit. By systematically evaluating different sampling frequencies, the proposed approach achieves improved response times and tracking accuracy.
\item Empirical validation of the proposed system under diverse channel conditions and operational scenarios. Performance evaluations demonstrate that the system consistently outperforms conventional tracking-based approaches in terms of both interception success rate and tracking efficiency.
\end{itemize}
By integrating swarm intelligence with asynchronous DRL, the proposed system transforms UAV-based security operations from passive monitoring into proactive intervention, significantly enhancing airspace safety and security

\begin{figure}[!t]
\centering
\includegraphics[width=8.5cm]{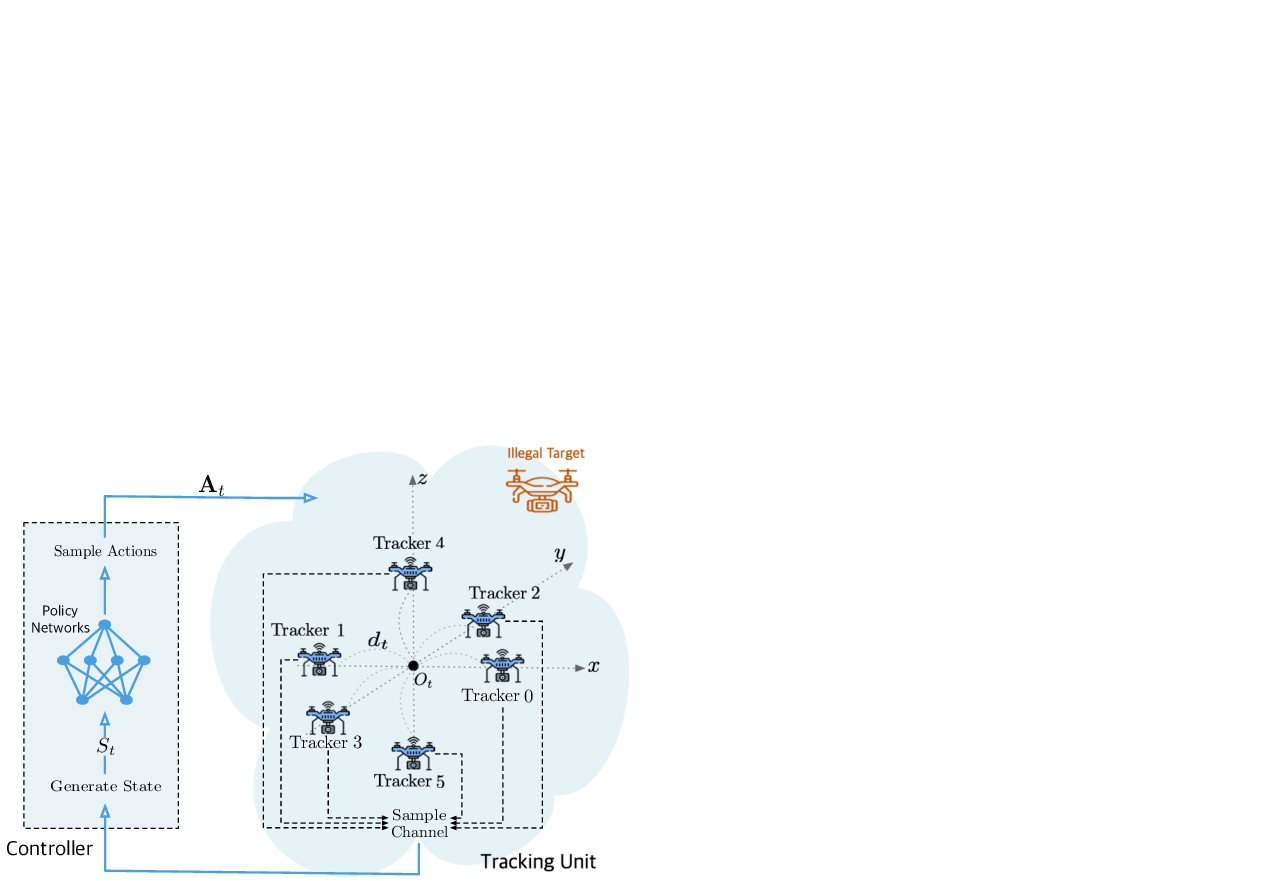}
\caption{Proposed system based on swarm flight using deep reinforcement learning for chasing illegal UAVs.}
\label{system_model} 
\end{figure}

\subsection{Paper Organization}
This paper is structured as follows. Section II describes the system architecture, including swarm flight methodologies and deep reinforcement learning techniques. Section III details the control framework, emphasizing UAV swarm coordination strategies. Section IV presents numerical results evaluating the system’s performance across varying operational conditions. Finally, Section V concludes the paper by summarizing key findings and discussing future research directions.

\section{System Model} \label{sec:2}
\subsection{Overall System Architecture}
In this paper, we introduce a system for chasing unauthorized UAVs based on swarm flight using deep reinforcement learning. The system comprises a controller and a tracking unit consisting of six compact UAVs, as illustrated in Fig. \ref{system_model}. The six tracking UAVs are positioned in three-dimensional space with two UAVs along each of the $x$, $y$, and $z$ axes. Each UAV  is denoted as $u,~0\le u \le 5$. At a specific time $t$, the center point of the tracking UAVs is denoted as $O_t$ and its coordinates are $(x_t, y_t, z_t)$. The distance between $O_t$ and each UAV is represented as $d_t$.

The tracking UAVs intercept communication signals that are exchanged between an illegal target UAV and its controller that refers to the entity responsible for operating the unauthorized UAV. This can typically be a ground control station (GCS), a remote pilot using a handheld transmitter, or a network-based control system issuing commands through a wireless communication link. The tracking UAVs then measure RSSI values at a given sampling frequency $F$ [samples/sec].
Subsequently, the RSSI values are transmitted to the controller through a cellular network or wireless local area network (WLAN). The controller utilizes the RSSI values to adjust both the movement of the tracking UAVs and the spacing between the tracking UAVs.

If the tracking unit is equipped with multiple RF antennas on a single UAV for system simplicity, the increased spatial correlation among antennas could lead to reduced diversity in the RSSI values measured by each antenna \cite{3gpp_ts}.
Consequently, the unit's ability to accurately predict the direction of the target may be compromised, impairing chasing performance.
Furthermore, such a configuration might struggle to operate reliably in three-dimensional spaces due to challenges in multidimensional arrangement.
Given these considerations, we decided to configure the tracking unit with multiple compact UAVs, each equipped with a single RF receiver, accepting increased system complexity as a trade-off for enhanced performance and reliability, particularly in three-dimensional spaces.

The controller utilizes a policy-gradient deep reinforcement learning approach, specifically the asynchronous advantage actor-critic (A3C) model, to manage the movements of the tracking unit. This is achieved by leveraging the RSSI values received from the tracking UAVs. Initially, the control unit generates the input state \( S_t \) based on these RSSI values. Subsequently, \( S_t \) is input into the policy networks, which stochastically determine the optimal spacing between the tracking UAVs as well as the tracking unit's movements along the \( x \), \( y \), and \( z \) axes. This setup enhances the tracking precision and response to dynamic target movements.

\subsection{Channel Model}
We posit that an unauthorized UAV communicates with its controller via a wireless network. For a given distance \(d\) between such an illegal UAV and a tracking UAV, the RSSI value measured at the tracking UAV is defined as
\begin{equation}
\gamma(d) = P_{\text{tx}} - L(d) + \xi~ [\text{dBm}].
\label{eq_rssi}
\end{equation}
Here, \(P_{\text{tx}}\) represents the transmit power utilized by the unauthorized UAV for communication, and \(L(d)\) is the path loss corresponding to distance \(d\). The path loss \(L(d)\) is further detailed as
\begin{equation}
L(d) = L_0 + 10n\log_{10}\left(\frac{d}{d_0}\right) [\text{dB}],
\end{equation}
where \(d_0\), \(L_0\), and \(n\) denote the reference distance, the path loss at that reference distance, and the path loss exponent, respectively. The term \(\xi\) in \eqref{eq_rssi} denotes small-scale fading. In this paper, we consider Rician fading, where one of the multipath components, typically a line-of-sight (LOS) signal, is much stronger than the others. This implies that \(\xi\) follows a Rician distribution with a \(K\)-factor, which quantifies the ratio of the LOS power to the total power of all indirect paths. Consequently, \(\xi\) is modeled to follow a Rician distribution with a \(K\)-factor, indicating the ratio of LOS power to the cumulative power of all indirect paths.

In this paper, we thoroughly explore the complex dynamics of wireless communication channels. Initially, we employ a correlation matrix framework to quantify the spatial correlation among antennas mounted on each tracking UAV. Spatial correlation characterizes the interdependence of signal strengths received at different locations, which primarily arises due to environmental factors such as terrain and the specific configuration of antenna deployment \cite{Brown2012}.
In this context, we define the spatial correlation matrix \(\mathbf{C}\) as \cite{matlab_spatial}
\begin{equation}
\mathbf{C} = \left[c_{ij}\right]_{0 \leq i, j \leq 5},
\end{equation}
where \(c_{ij}\) represents the correlation coefficient between two tracking UAVs \(i\) and \(j\), defined by
\begin{equation}
c_{ij} = 
\begin{cases} 
1 & \text{if } i = j, \\
\rho & \text{otherwise}.
\end{cases}
\end{equation}
Unlike traditional spatial correlation matrices that exhibit varying correlation coefficients between antenna elements, our approach maintains a uniform correlation coefficient \(\rho\) across all tracking UAVs. This uniformity stems from the symmetrical structure of the tracking UAVs, which resemble a square pyramid.

Furthermore, we enhance the realism of our channel model by accounting for Doppler shift, which affects channel behavior over time. Doppler shift, represented by \(f_d\), is defined as
\begin{equation}
f_d = \frac{v}{c} f_c,
\label{eq_doppler}
\end{equation}
where \(v\), \(c\), and \(f_c\) denote the relative velocity of tracking UAVs, the speed of light, and the carrier frequency, respectively \cite{Rappaport2002, matlab_temporal}.

\begin{figure*}[!t]
\centering
\includegraphics[width=14cm]{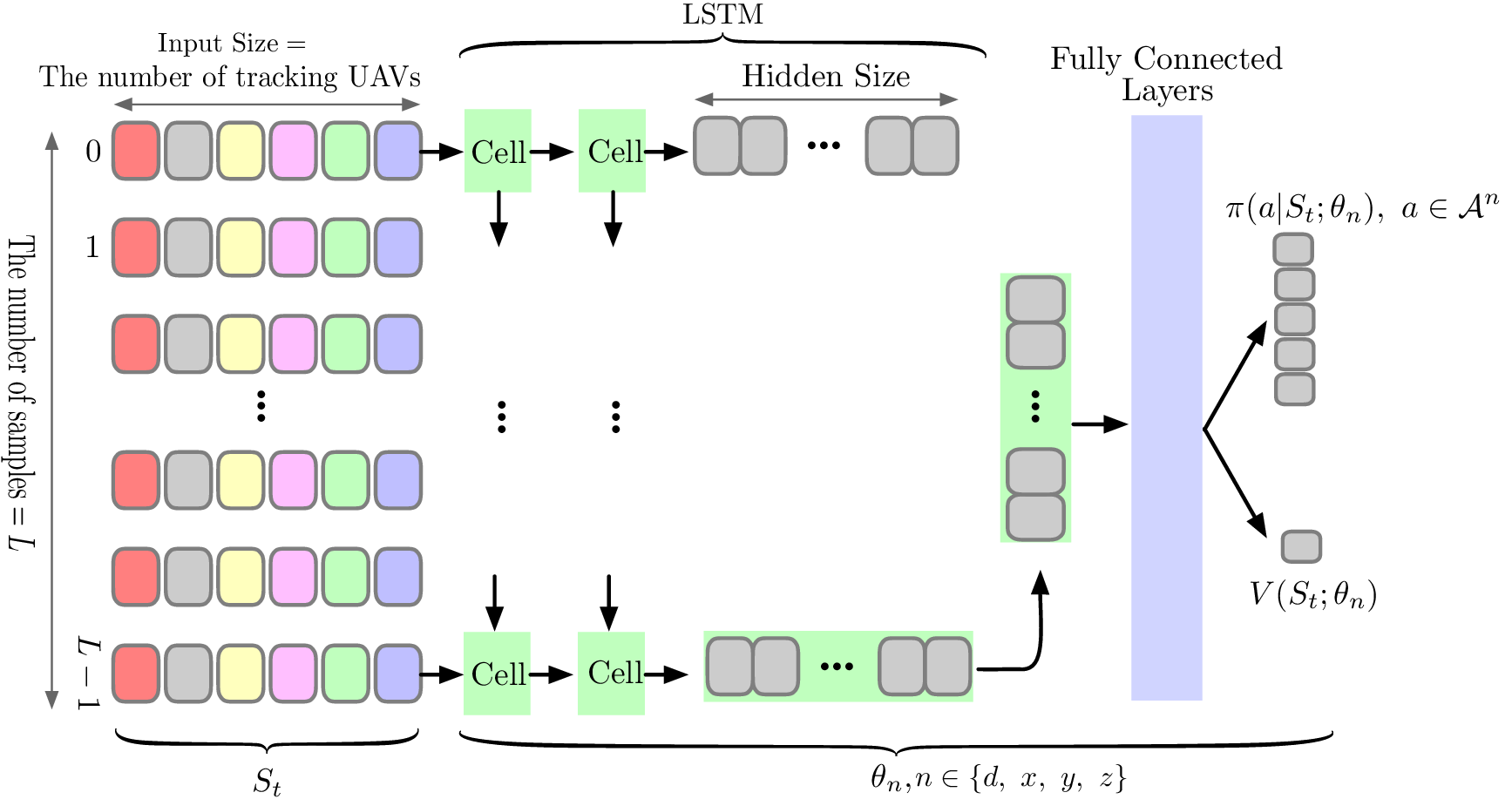}
\caption{The architecture of the proposed neural network.}
\label{a3c_neural_network}
\end{figure*}

\section{Control Framework and Asynchronous Training Using Tracking Controller}\label{sect_proposed}

\subsection{Forward Pipeline For The Control Framework}\label{sect_pipeline}
As outlined in Section II, this paper adopts the A3C model, leveraging its significant benefits, including asynchronous training that boosts learning speed and stability through the use of multiple agents.
It combines value-based and policy-based methods to develop robust policies, effectively manages high-dimensional state and action spaces with deep neural networks, and exhibits reduced sensitivity to hyper-parameter settings.
These features make the A3C model widely applicable and efficient across diverse tasks and environments \cite{mniha16}.

The controller proposed in this study, utilizing the A3C model as depicted in Figure \ref{system_model}, is equipped with four distinct policy networks.
Each of the four policy networks is uniquely parametrized by $\theta^n$, specializing in different tracking dimensions where $n \in {d, x, y, z}$.
$\theta^d$ adjusts the spacing between the tracking UAVs, as indicated by $d_t$ in Figure \ref{system_model}, to ensure optimal positioning for pursuing an unauthorized target. 
Meanwhile, $\theta^x$, $\theta^y$, and $\theta^z$ manage the movements of the tracking unit along the $x$, $y$, and $z$ axes, respectively, enabling precise maneuverability in three-dimensional space.

\begin{itemize}
\item \textbf{State}: The input state $S_t$ at any given time $t$ is constructed using the RSSI values collected by each tracking UAV. The matrix representation of $S_t$ is defined as
\begin{align}
S_t = \left[ \gamma_{iu} \right]_{0 \leq i \leq L-1,~0 \leq u \leq 5},
\end{align}
where $\gamma_{iu}$ represents the $i$-th RSSI sample measured by the $u$-th UAV. The variable $L$ indicates the total number of RSSI samples measured by each UAV within this interval, and the UAV index $u$ runs from 0 to 5, assuming there are six tracking UAVs in the tracking unit as depicted in Fig. \ref{system_model}.

Each RSSI sample $\gamma_{iu}$ is obtained at a sampling frequency $F$ [samples/sec], which determines how often RSSI measurements are recorded within the given interval. The arrangement of these RSSI values in matrix is depicted in Fig. \ref{a3c_neural_network}. This formulation implies that each row in matrix $S_t$ corresponds to the synchronized RSSI samples taken by all UAVs at a single sampling instance, while each column represents a series of RSSI samples from a single UAV across all sampling instances.

\item \textbf{Policy and value functions}: As depicted in Fig. \ref{a3c_neural_network}, input state $S_t$ is processed by four separate neural networks, which consist of long short-term memory (LSTM) layers and fully connected layers. 
LSTMs are particularly well-suited for tasks where understanding the temporal dynamics of input sequences is crucial. They are capable of learning long-term dependencies in the data, avoiding the vanishing gradient problem common in standard recurrent neural networks.
Following the LSTM layers, the final hidden state of the LSTM layers is then passed through fully connected (dense) layers. Each of these fully connected layers uses the Rectified Linear Unit (ReLU) activation function.

Each network produces two critical outputs: a policy  $\pi(a|S_t;\theta^n)$ and a value function $V(S_t;\theta^n)$. The policy output specifies the probability distribution across possible actions, directing the agent's decision-making by indicating the most favorable action based on the current state $S_t$.
Simultaneously, the value output provides an estimation of the expected cumulative reward from taking a specific action in the given state, thereby informing the agent about potential future benefits. 
This design ensures that the agent remains adaptive and responsive to changes in the environment, enhancing its ability to make informed decisions that consider both present circumstances and future possibilities.
 
\item \textbf{Actions}: An action $A^{n}_{t}$ is stochastically sampled based on the policy $\pi(a|S_t;\theta^n)$. This action is drawn from its designated action space $\mathcal{A}^n$. Specifically, $\mathcal{A}^{d} = \{1, 2, 3, 4, 5\} [\text{m}]$ and $\mathcal{A}^{x}, \mathcal{A}^{y}, \mathcal{A}^{z}= \{-4, -2, 0, 2, 4\} [\text{m}]$.
In the proposed model, four action spaces are designed to be orthogonal to each other, underscoring the fact that actions generated by different policy networks do not interfere with each other. This orthogonality ensures that adjustments in one dimension do not inadvertently affect the performance in the other dimensions. As a result, the proposed tracking system can execute complex maneuvers with precision, as the implementation of one action does not necessitate compensatory adjustments in the others.

\end{itemize}

\subsection{Training of Proposed Model}

The tracking controller adjusts the positions of the tracking UAVs to carry out a determined set of actions, $\mathbf{A}_{t}=[A^d_{t}, A^x_{t}, A^y_{t}, A^z_{t}]$. After implementing these actions, the controller observes the subsequent state $S_{t+1}$ and collects associated rewards, $\mathbf{R}_{t+1}=[R^d_{t+1},R^x_{t+1},R^y_{t+1},R^z_{t+1}]$. Utilizing these observation and rewards, the controller then computes losses and updates the parameters of the neural networks to minimize the losses.

\begin{figure*}[!t]
\centering
\includegraphics[scale=0.8]{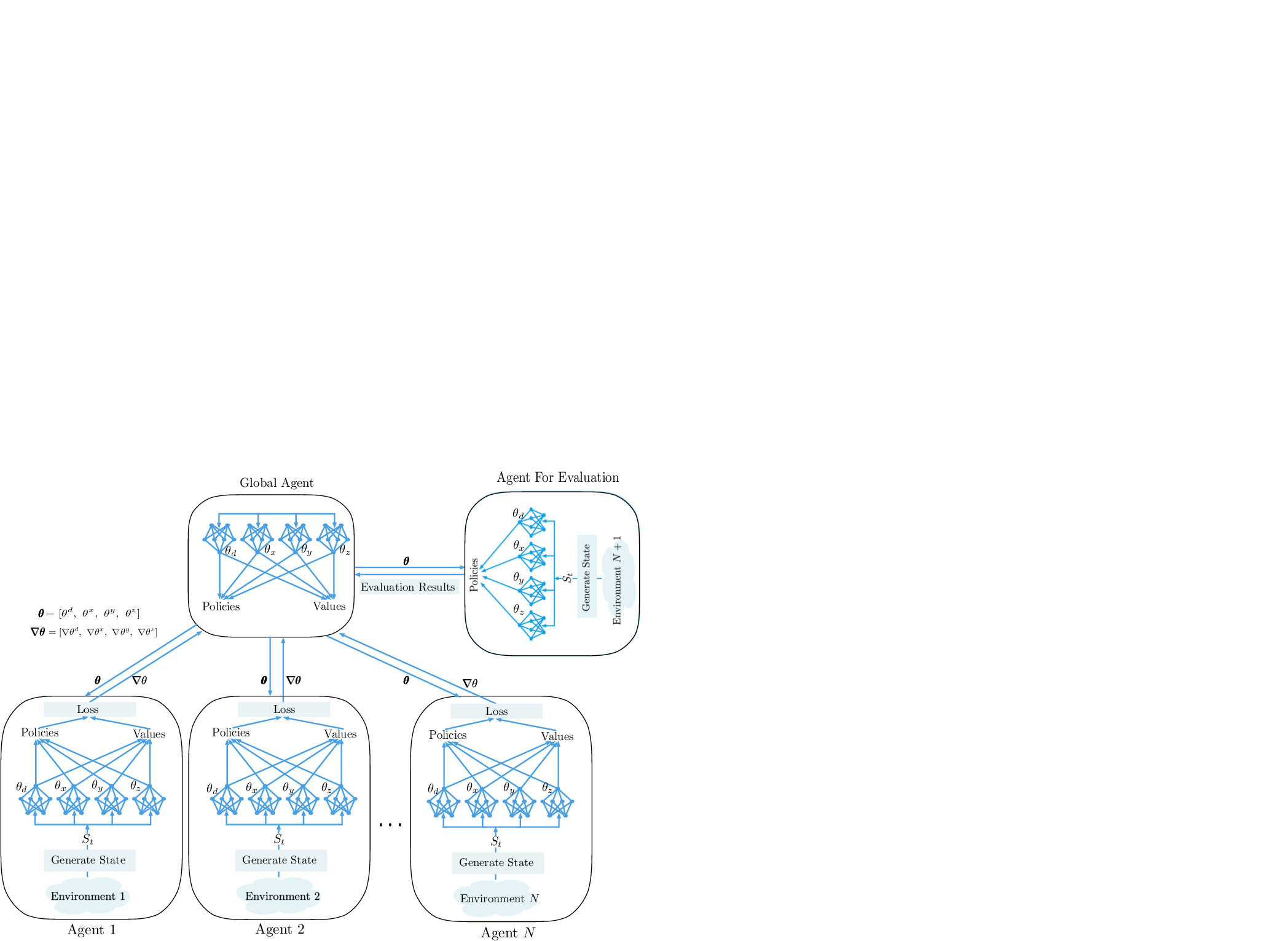}
\caption{Training architecture using shared learning and asynchronous update for the proposed model }
\label{a3c_architecture} 
\end{figure*}

 \begin{itemize}
\item \textbf{Rewards}: In this study, we have developed a specialized reward structure tailored for four distinct neural networks, each designed to generate orthogonal actions within a multi-agent deep reinforcement learning framework. This unique reward system is critical for aligning the learning process with the individual objectives of each policy network, thereby enhancing the tracking performance of our UAV chasing system.

For each axis-aligned action \(A_t^n\) where \(n \in \{x, y, z\}\), the reward \(R_{t+1}^n\) is computed based on the decrease in the absolute difference in the corresponding-axis coordinates between the tracker's center point and the target UAV. Specifically, the reward is calculated as
\begin{align}
R_{t+1}^n = \left| O_t[n] - U[n] \right| - \left| O_{t+1}[n] - U[n] \right|
\end{align}
for each \(n\) axis coordinate. Additionally, the reward for the action \(A_t^d\) is defined as the reduction in the Euclidean distance between the tracker and the target UAV:
\begin{align}
R_{t+1}^b = \left| O_t - U \right| - \left| O_{t+1} - U \right|
\end{align}
where \(\left|O_t - U\right|\) denotes the Euclidean distance between the tracking center point \(O_t\) and the target UAV \(U\).

These tailored rewards ensure that the learning outcomes are directly tied to the effectiveness of the UAV's tracking ability, promoting a more efficient and targeted learning process across different control dimensions.

\item \textbf{Losses} In this study, the loss function is defined with three primary components: policy loss, value loss, and an entropy bonus. These components are strategically combined to optimize the policy and value function networks across the update interval from $t$ to $t+T-1$.
Each neural network, indexed by \( n \) from the set \(\{d, x, y, z\}\) and parametrized by \(\theta^n\), processes a sequence of \(T\) observations \(\left(\! (S_t,\! A_t^n, \!R_{t+1}^n, \!S_{t+1}), \!\ldots, \!(S_{t+T-1}, \!A_{t+T-1}^n,\! R_{t+T}^n, \!S_{t+T}) \!\right)\) over \( T \) time steps starting from time $t$.

The policy loss \( L_P^n \) is formulated to encourage actions that are expected to yield higher returns. It utilizes the advantage function, which measures the relative benefit of selected actions against a baseline expectation. This is accomplished by calculating the difference between the temporal-difference (TD) target and the current value estimate. The advantage function at time \( t \) is defined as
\begin{align}
ADV_t^n = G^n_t - V(S_t; \theta^n),
\label{eq_adv}
\end{align}
where \( G^n_t \) represents the TD target at state \( S_t \), and \( V(S_t; \theta^n) \) is the estimated value of the current state. Typically, \( G^n_t \) is calculated using the value of the subsequent state \( V(S_{t+1}; \theta^n) \) combined with the immediate reward, but in this study, we compute it using the discounted sum of future rewards
\begin{align}
G^n_t = \sum_{k=1}^{T-t+1} \gamma^{k-1} R^n_{t+k},
\end{align}
where \( \gamma \) is the discount factor, prioritizing immediate rewards and diminishing the importance of future rewards. If the final state is terminal, \( R^n_{T+1} = 0 \); otherwise, \( R^n_{T+1} = V(S_T; \theta^n) \), indicating the absence of future rewards beyond the terminal state.
This method of calculating \( G^n_t \), which incorporates a series of future rewards, provides a comprehensive assessment of potential outcomes, enhancing the robustness and accuracy of the advantage function. This approach is particularly effective when \( T \), the length of the trajectory, is long. It allows the system to account for the extended consequences of actions, ensuring that the advantage function reflects the cumulative impact over the entire trajectory.

The policy loss is subsequently calculated as
\begin{align}
L_P^n = -\frac{1}{T} \sum_{t=0}^{T-1} \log \pi(A_t^n \mid S_t; \theta^n) \cdot ADV_t^n.
\label{eq_policy_loss}
\end{align}

The value loss \( L_V^n \) is crucial for aligning the predictions of the value function with the actual returns observed. To improve model stability and accuracy, the value loss is calculated using a TD approach that directly leverages the next-step value predictions to form a more responsive learning update.  \( L_V^n \) is defined using the absolute difference (L1-loss) between the predicted value at a state and the temporal-difference target. This can be mathematically expressed as
\begin{align}
L_V^n = \frac{1}{T} \sum_{t=0}^{T-1} \left|  R_{t+1}^n + \gamma V(S_{t+1}; \theta^n) -V(S_t; \theta^n) \right|,
\end{align}
 where $V(S_{T}; \theta^n)=0$ if \( S_T \) is a terminal state.
 
 To encourage exploration and prevent premature convergence to suboptimal policies, an entropy bonus, denoted as $H^n$, is incorporated into the loss calculation. This entropy bonus measures the uncertainty of the policy distribution at each time step and is calculated as
\begin{equation}
H^n=-\frac{1}{T}\sum\limits_{t=0}^{T-1}\sum_{a\in \mathcal{A}^n}\!\pi(a|S_t;\theta^n)\!\log \pi(a|S_t;\theta^n)).
\end{equation}

Combining all these components, the total loss over an update interval \( T \) is formulated as
\begin{equation}
    L^n = L^n_P + L^n_V -\beta H^n,
\label{eq_loss}
\end{equation}
where $\beta$ is an coefficient that adjust the relative importance of entropy bonus and $H^n$ is subtracted from the loss to promote a more exploratory and diverse policy behavior.

\item \textbf{Shared learning and asynchronous update}: Based on \eqref{eq_loss}, each agent calculates the gradients of local parameters for each neural network ${n\in\{d,x,y,z\}}$ with respect to $L^n$ , denoted as $\boldsymbol{\nabla\theta}= [\nabla{\theta^n}]$. These gradients are then transmitted to a central or global agent. The global agent updates the networks asynchronously, incorporating the gradients derived from local agents' experiences. This method of shared learning accelerates the enhancement of policy and value estimates across all agents by leveraging their collective experiences. Moreover, the asynchronous update mechanism eliminates the requirement for agents to synchronize their learning phases, thereby enabling more continuous and efficient training cycles.

\item \textbf{Shared Learning and Asynchronous Update}: Following the loss formulation in \eqref{eq_loss}, each agent independently computes the gradients for their local parameters of four neural networks, denoted by $\boldsymbol{\nabla\theta}= \{\nabla{\theta^n}|n \in \{d, x, y, z\}\}$. These gradients are then transmitted to the global agent. The global agent updates the network parameters asynchronously, integrating the gradients collected from the experiences of all local agents. This shared learning approach significantly accelerates the improvement of policy and value function estimates by utilizing the collective insights gained from multiple agents. Additionally, the asynchronous update mechanism facilitates continuous and efficient training cycles by removing the need for synchronization among the agents, thus enhancing the overall scalability and responsiveness of the learning process.
\end{itemize}

The overall architecture and training algorithm for the proposed model are depicted in Fig. \ref{a3c_architecture} and detailed in Algorithm 1, respectively. This framework integrates \(N\) local agents and one global agent, all operating in a multi-threaded or multi-processor environment to enable shared learning and asynchronous updates effectively. An additional auxiliary agent is incorporated to assess intermediate training outcomes periodically. Should these outcomes meet predefined performance criteria, training may be terminated early, thereby enhancing the efficiency and effectiveness of the process.

\begin{algorithm}[t]
\label{algo_training}
\caption{Training algorithm of the proposed model}
\SetAlgoLined
\SetAlgoNoEnd
\DontPrintSemicolon
Generate local networks

\For{$ep=0$ \KwTo EPISODES}{
    Initialize environment and state $S_t$\;
    $done \leftarrow \text{False}$\;
    $cnt=0$\;
    \While{not done}{
        Clear memory\;
	    Synchronize local networks with global networks\;        
        \For{$t = 0$ \KwTo $T-1$}{
            \ForAll{$n \in \{d,x,y,z\}$}{
                Select $A^n_{t}$ from $\pi(a|S_t;\theta^n)$\;
            }
            $\mathbf{A}_{t}=[A^d_{t}, A^x_{t}, A^y_{t}, A^z_{t}]$\;
            Execute $\mathbf{A}_{t}$\;
            Observe $\mathbf{R}_{t+1}$ and $S_{t+1}$\;
            Store $(S_t, \mathbf{A}_{t}, \mathbf{R}_{t+1}, S_{t+1})$\;
            $S_t \leftarrow S_{t+1}$\;
    	    \If{$cnt\ge$ MAX\_STEPS}{
				$done \leftarrow \text{True}$\tcp*{Failure}
	        }
    	    \If{$| O_{t+1}-U|\le th$}{
				$done \leftarrow \text{True}$  \tcp*{Success}
	        }
            
            \If{done}{
                break\;
            }
	        $cnt=cnt+1$\;            
        }
        \ForAll{$n \in \{d,x,y,z\}$}{
            $G^n_{t+1}\!\!=\!\!0$ if terminal, else $G^n_{t+1}\!\!=\!\!V\!(\!S_{t+1}\!;\!\theta^n\!)$\;
            $L^n_P=0$;$L^n_V=0$;$H^n=0$\;
            \For{$k = t$ \KwTo 0}{
                Calculate the discounted sum of rewards and advantage:\;
                $G^n_k = R^n_{k+1} + \gamma G^n_{k+1}$\;
                $ADV^n_k = G^n_k - V(S_k;\theta^n)$\;
                Calculate the policy (actor) loss:\;
                $L^n_P = L^n_P -ADV^n_k \cdot \log \pi(A^n_{k} | S_k;\theta^n)$\;
                Calculate the critic loss:\;
                $L^n_V = L^n_V + |ADV_k|$\;
                Calculate the entropy:\;
                $H^n\!\! =\!\! H^n\!-\!\sum_A \pi_{\theta_a}(A | S_k)\!\!\log(\pi_{\theta_a}(A | S_k))$\;
            }
            Calculate the overall loss:\;
            $L^n = \frac{1}{t+1} [L^n_P + L^n_V - \beta H^n]$\;
            Calculate the gradients $\nabla{\theta_a}(L^n)$\;
        }
        Transfer $[\!\nabla{\theta^d}, \!\!\nabla{\theta^x}, \!\!\nabla{\theta^y}, \!\!\nabla{\theta^z}\!]$ to the global agent\;
        \ForAll{$n \in \{d,x,y,z\}$}{
            $\theta^n \gets \theta^n - lr \cdot \nabla{\theta^n}$ \tcp*{Global agent}
        }
    }
}
\end{algorithm}

\section{Numerical Results and Discussion} \label{sec_numerial_results}

\begin{table}[t]
\centering
\caption{Parameters}
\label{table_parameters}
\begin{tabular}{ll|r}
\hline
\multicolumn{2}{c|}{Parameters}                                                          & Values \\ \hline\hline
\multicolumn{1}{c|}{$P_{\text{tx}}$} & Transmit power of target UAV      & 23 dBm \\\hline
\multicolumn{1}{c|}{$d_0$}           & Reference distance of path loss   & 1 m    \\\hline
\multicolumn{1}{c|}{$L_0$}           & Reference path loss at $d_0$      & 30 dB  \\\hline
\multicolumn{1}{c|}{$n$}             & Path loss exponent                & 2.6    \\ \hline
\multicolumn{1}{c|}{$K$}             & $K$-factor of Rician distribution & 3      \\ \hline
\multicolumn{1}{c|}{$f_d$}           & Doppler spread                    & 1 Hz   \\ \hline
\multicolumn{1}{c|}{$L$}             & The length of input states $S_t$  & 50     \\\hline
\multicolumn{1}{c|}{$T$}        & Update interval                & 5   \\\hline
\multicolumn{1}{c|}{$\gamma$}        & Discounting factor                & 0.99   \\\hline
\multicolumn{1}{c|}{$\beta$}         & Coefficient for entropy bonus     & 0.01   \\\hline
\multicolumn{1}{c|}{$th$} & \begin{tabular}[c]{@{}l@{}}Min. required distance btw tracking\\ and target UAVs required for success\end{tabular} & 2 m \\\hline
MAX\_STEPS      & Max. steps allowable for success   & 500    \\\hline
\multicolumn{2}{l|}{Number of LSTM layers}                               & 3      \\\hline
\multicolumn{2}{l|}{Number of features in the hidden state of LSTM}      & 128    \\\hline
\multicolumn{2}{l|}{Number of linear layers}                             & 2      \\\hline
\multicolumn{2}{l|}{Number of features of linear layers}                 & 128    \\\hline
\multicolumn{2}{l|}{Optimizer}                                           & Adam   \\\hline
\multicolumn{2}{l|}{Learning rate}                                       & $10^{-5}$   \\\hline
\multicolumn{2}{l|}{Number of local agents}                              & 8   \\\hline
\hline
\end{tabular}
\end{table}

In this section, we evaluate the performance of the UAV chasing system, as described in Sect. \ref{sect_proposed}. The analysis considers various channel sampling frequencies and spatial correlation coefficients, denoted as \(F\) and \(\rho\), respectively. The target UAV is positioned at a distance of 100 meters from the tracking system's center and is assumed to be randomly located on the surface of a sphere with a radius of 100 meters, ensuring \(|O_0 - U| = 100\) m.

Key parameters related to the channel model, neural networks, and training processes are detailed in Table \ref{table_parameters}. The relative velocity of the tracking UAVs is set at 0.1 m/s, while the frequency emitted by the target UAV is 3 GHz. The resulting Doppler shift, calculated according to \eqref{eq_doppler}, is 1 Hz.

\begin{figure}
     \centering
     \begin{subfigure}[b]{0.5\textwidth}
         \centering
         \includegraphics[width=8.5cm]{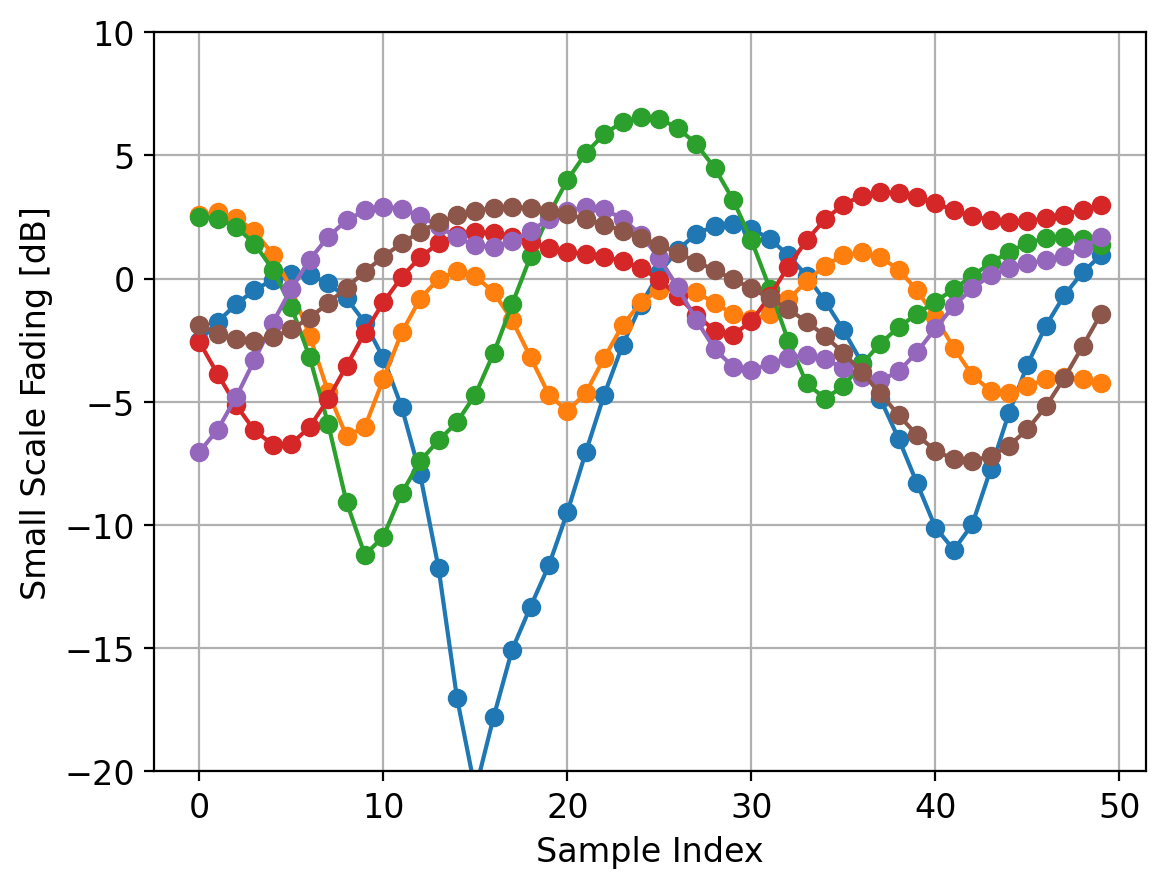}
         \caption{$F=20$Hz and $\rho=0.1$}
     \end{subfigure}
     \hfill
     \begin{subfigure}[b]{0.5\textwidth}
         \centering
         \includegraphics[width=8.5cm]{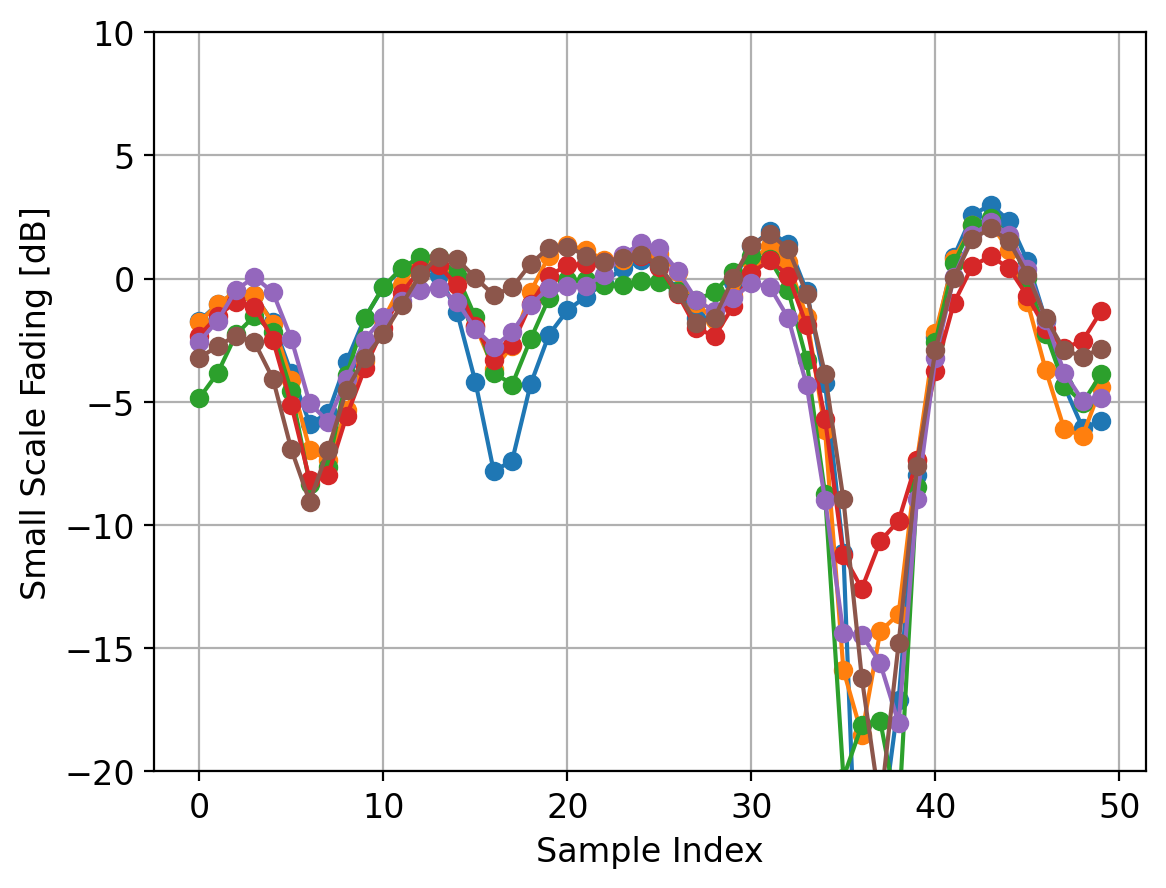}
         \caption{$F=10$Hz and $\rho=0.9$}
     \end{subfigure}
        \caption{Examples of small scale fading of six channels between each tracking UAV and target UAV when $f_d=1$ Hz.}
        \label{fig_ch_fading}
\end{figure}

Fig. \ref{fig_ch_fading} presents examples of small-scale fading in the channels between each of the six tracking UAVs and a target UAV, under a Doppler shift condition of \(f_d=1\) Hz. The influence of channel sampling frequency, \(F\), on the temporal resolution of the channel's characteristics is highlighted. As depicted in Fig. \ref{fig_ch_fading}-(a), a higher sampling frequency (\(F=20\) Hz) provides finer resolution, capturing detailed fluctuations in the fading process. Conversely, a lower sampling frequency (\(F=10\) Hz) as shown in Fig. \ref{fig_ch_fading}-(b), results in more pronounced variations between samples due to longer intervals, which can capture significant changes but at the cost of longer channel sampling periods. Furthermore, a low spatial correlation coefficient (\(\rho=0.1\) in Fig. \ref{fig_ch_fading}-(a)) indicates less correlation among the channel paths, suggesting a greater diversity in path characteristics. In contrast, a high spatial correlation coefficient (\(\rho=0.9\) in Fig. \ref{fig_ch_fading}-(b)) indicates highly correlated channel conditions, likely due to similar physical or environmental factors.

\begin{figure}
     \centering
         \includegraphics[width=8.5cm]{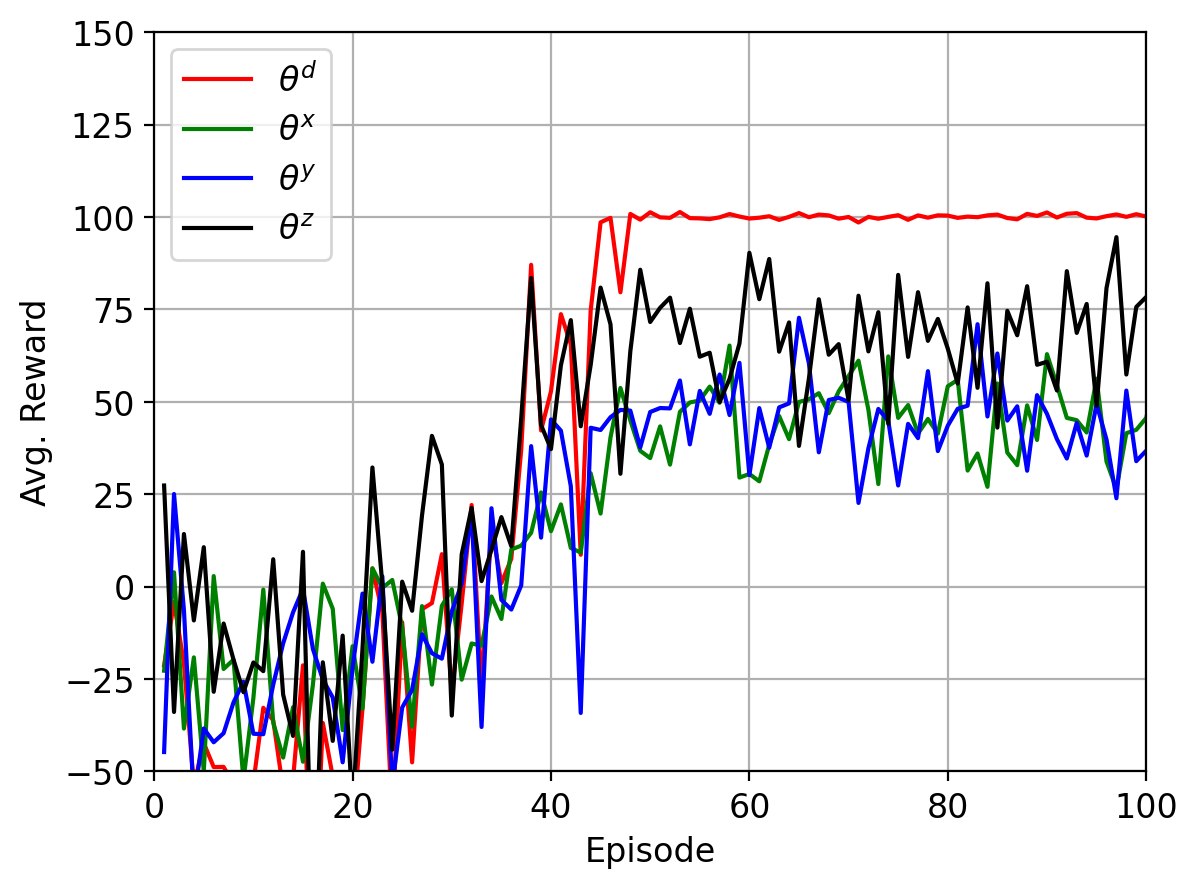}
         \caption{Average training reward when $F=100$ Hz and $\rho=0.5$.}
        \label{fig_train_reward}
\end{figure}

The proposed neural networks, denoted as \(\theta^n, n = \{d, x, y, z\}\), have been asynchronously trained by eight local agents, as detailed in Table \ref{table_parameters}. Figure \ref{fig_train_reward} displays the average rewards obtained over 100 episodes from these agents, illustrating a progressive improvement in all networks. Notably, the reward for \(\theta^d\) remains relatively stable and consistently higher compared to those for \(\theta^x\), \(\theta^y\), and \(\theta^z\).

The UAV tracking system's performance, utilizing these trained networks, was subsequently assessed over 10,000 episodes across varying channel sampling frequencies \(F\) [Hz] \(\in \{10, 20, 50, 100\}\) and spatial correlation coefficients \(\rho \in \{0.1, 0.5, 0.9\}\). This evaluation employed datasets distinct from those used in training, ensuring robust testing of the system's effectiveness.

\begin{figure*}
     \centering
     \begin{subfigure}[b]{0.32\textwidth}
         \centering
         \includegraphics[width=\textwidth]{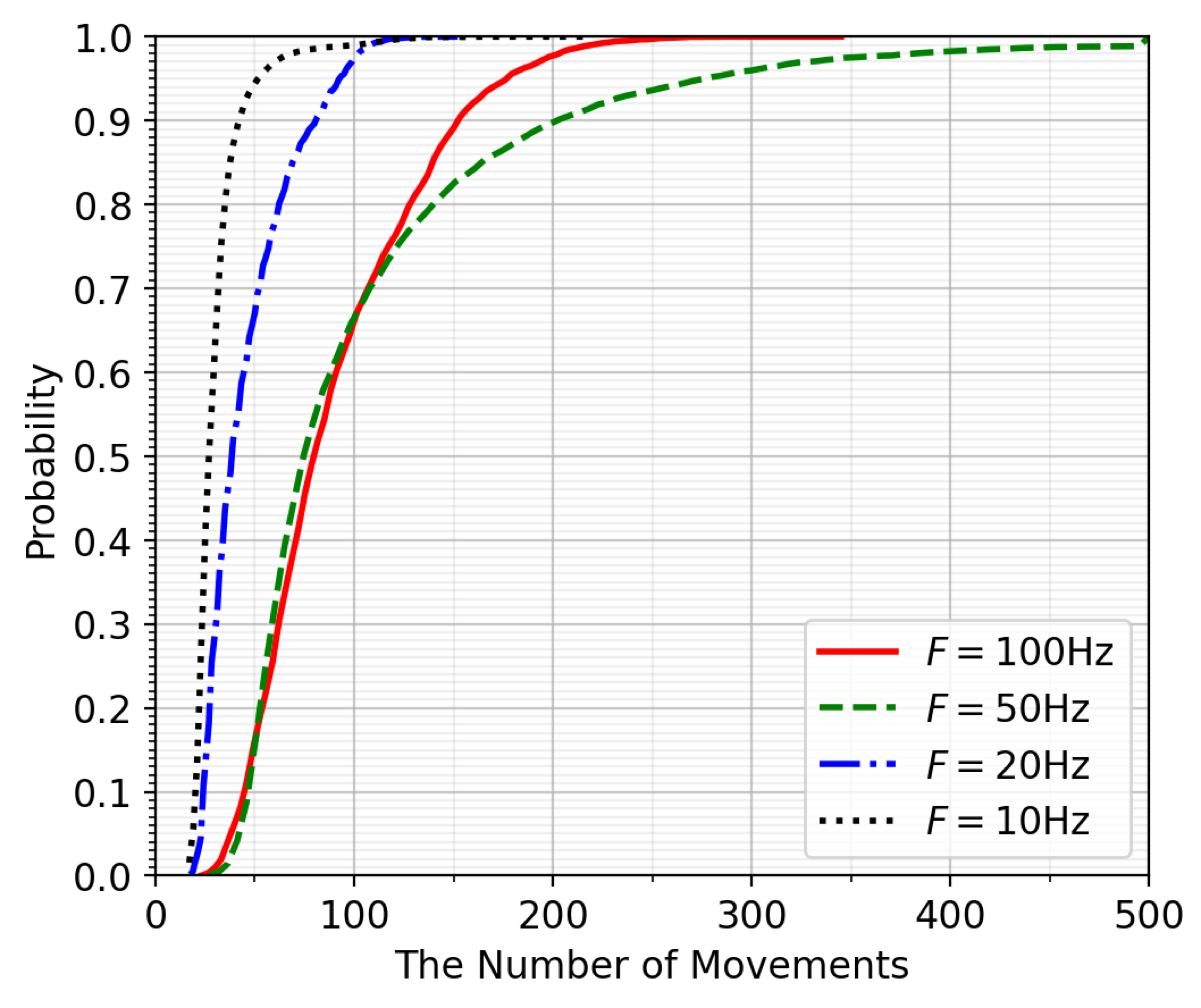}
         \caption{$\rho=0.1$}
     \end{subfigure}
     \begin{subfigure}[b]{0.32\textwidth}
         \centering
         \includegraphics[width=\textwidth]{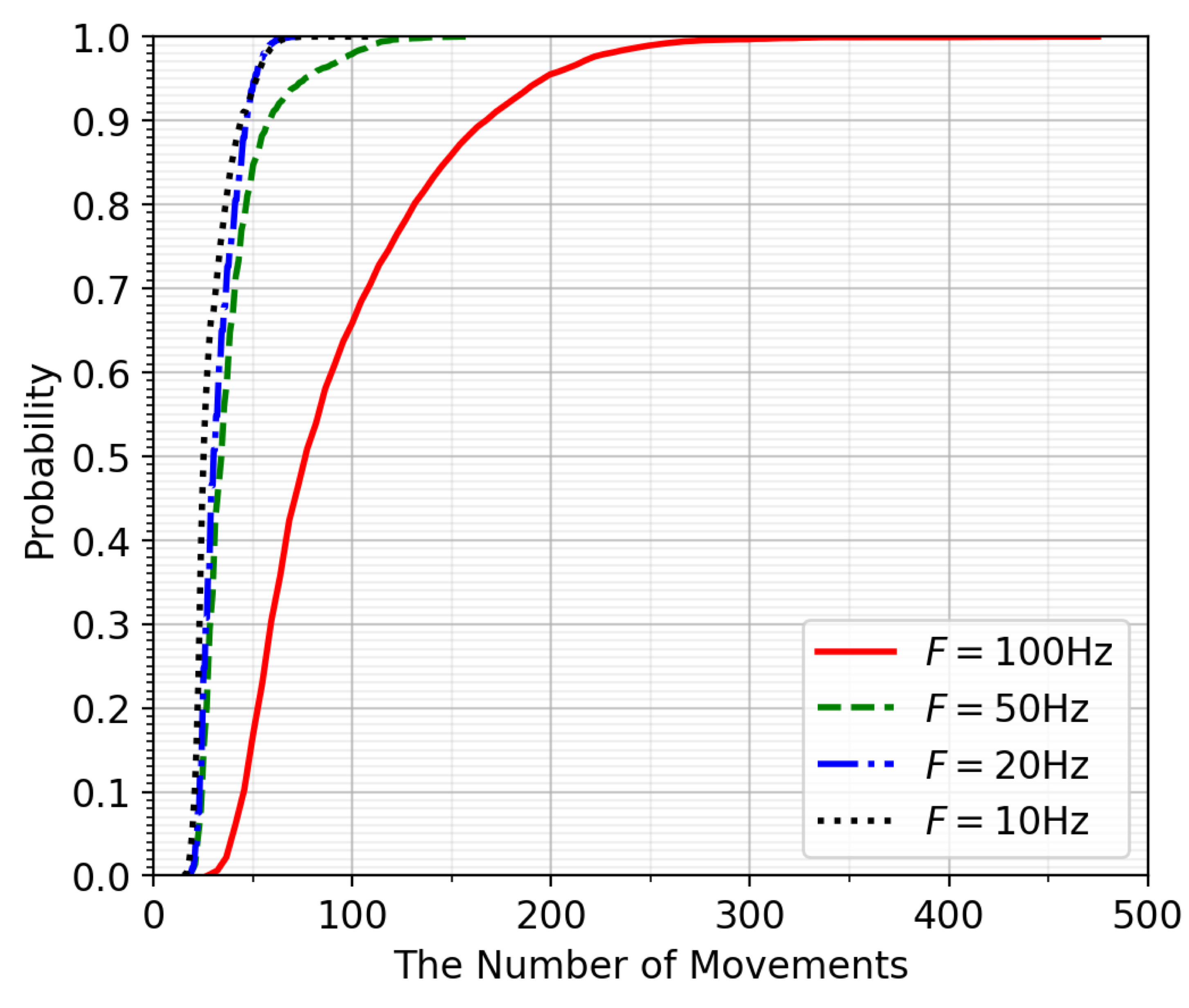}
         \caption{$\rho=0.5$}
     \end{subfigure}
     \begin{subfigure}[b]{0.32\textwidth}
         \centering
         \includegraphics[width=\textwidth]{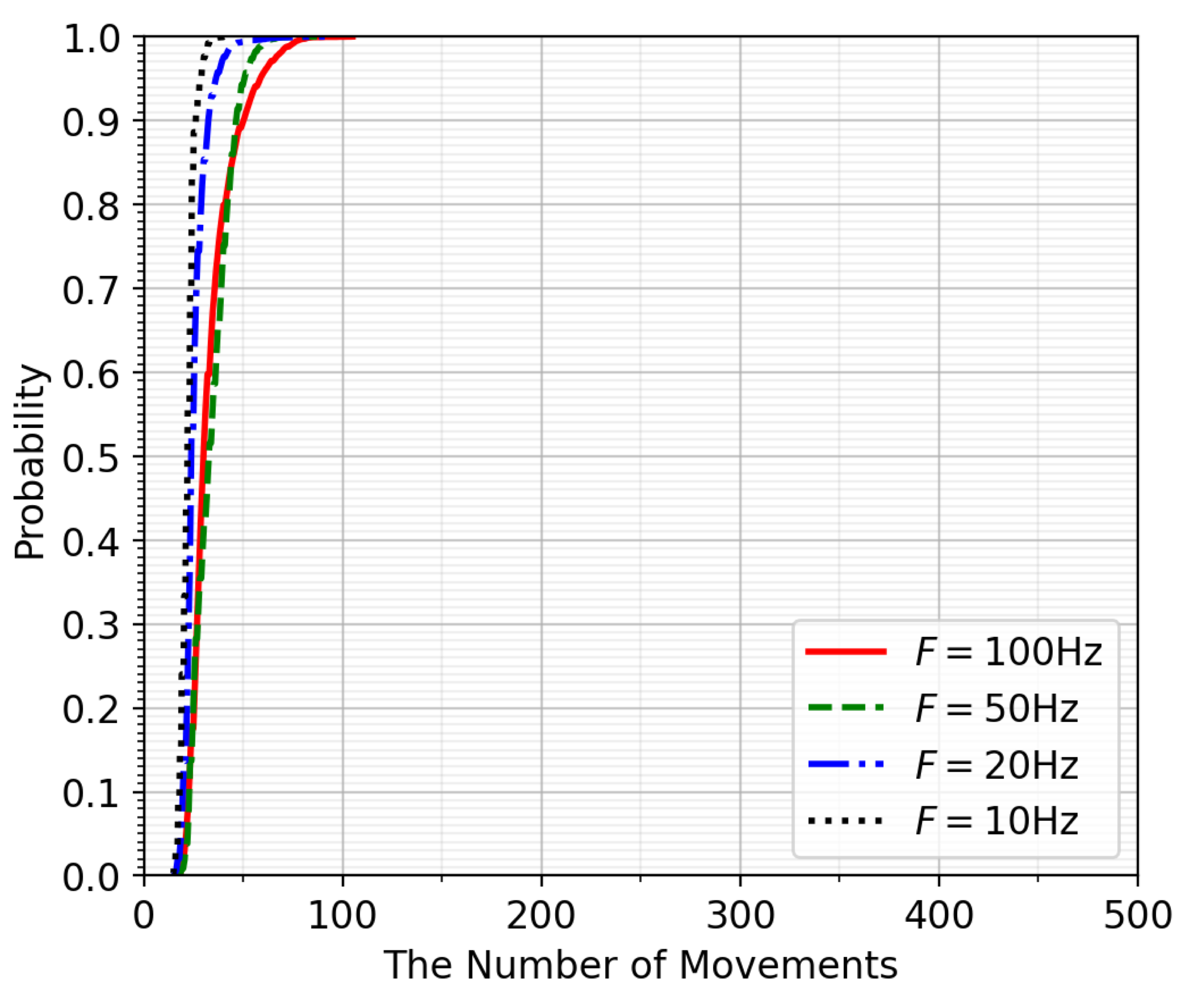}
         \caption{$\rho=0.9$}
     \end{subfigure}
        \caption{CDF of number of movements required to successfully track the target UAV for various spatial correlation coefficient $\rho\in\{0.1, 0.5, 0.9 \}$.}
        \label{fig_cdf_steps}
\end{figure*}

Figs. \ref{fig_cdf_steps} present the cumulative distribution functions (CDFs) for the number of movements required by the proposed chasing system to successfully chase an unauthorized target UAV within 2 meters. The system consistently achieves successful chasing within the criterion of fewer than 500 movements across various sampling frequencies and spatial correlation coefficients. This robust performance highlights the system's adaptability and reliability under diverse environmental and operational conditions.

As spatial correlation increases from $\rho=0.1$ in Fig. \ref{fig_cdf_steps}-(a) to $\rho=0.9$ in Fig. \ref{fig_cdf_steps}-(c), the CDF curves become noticeably steeper. A lower spatial correlation (\(\rho=0.1\)) suggests a less coherent channel environment, complicating the tracking process due to the increased complexity in received signals. In contrast, a higher spatial correlation (\(\rho=0.9\)) indicates more homogeneous channel conditions, simplifying the tracking process by reducing signal variability, thus enhancing tracking efficiency and reducing the required movements.

Regarding sampling frequency, higher frequencies result in closely spaced samples that may lead to data redundancy due to the similarity of information between samples, while lower frequencies extend the intervals between samples, increasing the diversity and utility of the information each sample provides. As illustrated in Figs. \ref{fig_cdf_steps} for all \(\rho\) values, reduced sampling frequencies effectively decrease the number of movements necessary for successful chasing, thereby optimizing the efficiency of the system.

\begin{figure}
     \centering
         \includegraphics[width=8.5cm]{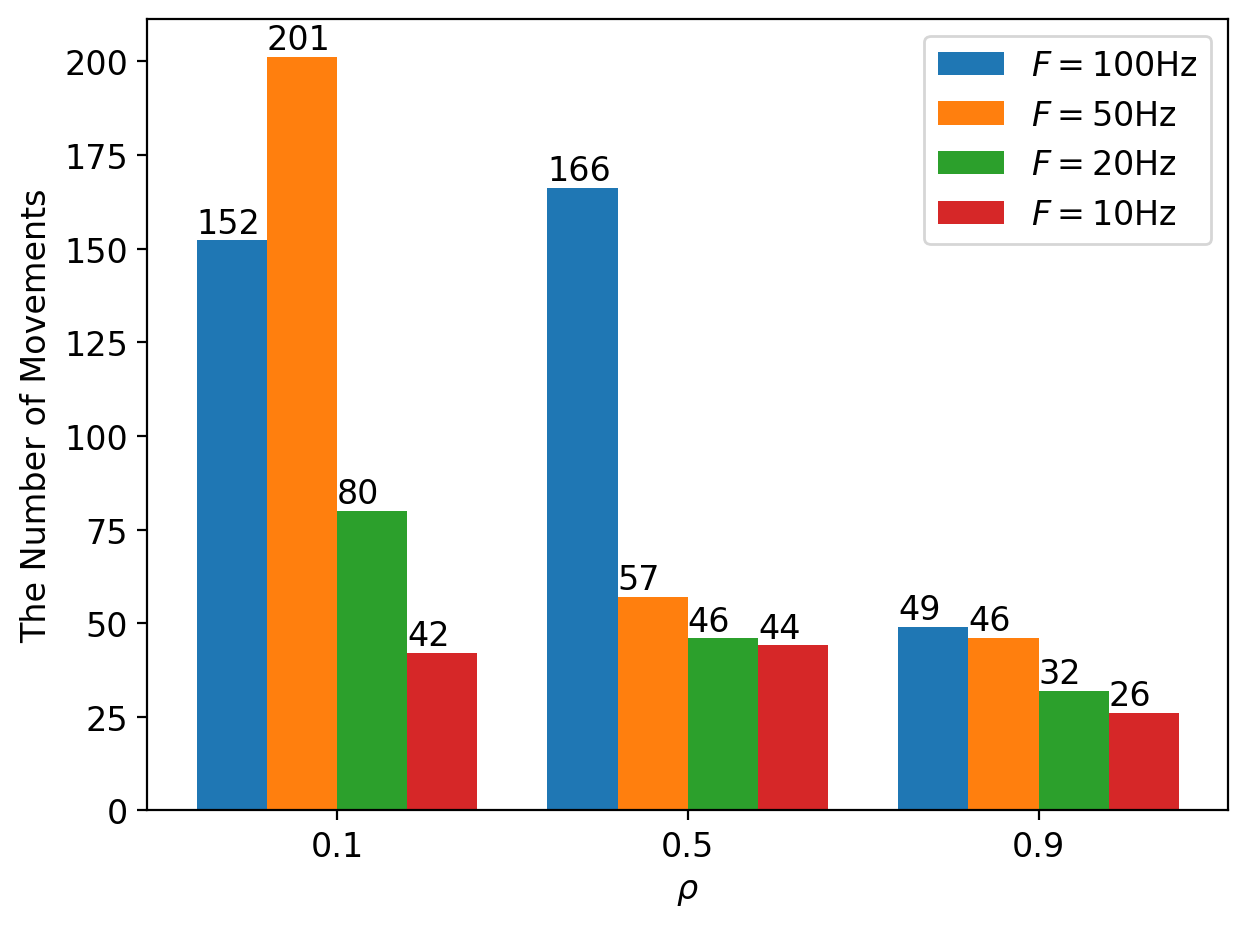}
         \caption{The number of movements for the 90th percentile.}
        \label{fig_step_percentile}
\end{figure}



While Figs. \ref{fig_cdf_steps} provide an overview of CDFs for the number of movements required to track and chase a target UAV, Fig. \ref{fig_step_percentile} specifically examines the 90th percentile. This analysis offers detailed insights into the system’s performance under near-worst-case scenarios, where the number of movements exceeds those of 90\% of chasing attempts. An increase in sampling frequency typically requires more movements at the 90th percentile across various spatial correlation coefficients due to less comprehensive channel information. For example, at $\rho=0.5$, increasing $F$ from 10Hz to 100Hz raises the number of movements from 44 to 160 at the 90th percentile. Interestingly, at the lowest spatial correlation ($\rho=0.1$), increasing $F$ to 100Hz results in fewer movements than at $F=50$Hz, indicating a complex relationship between sampling frequency and environmental dynamics.

For a given correlation coefficient \( \rho \), the number of movements required to successfully chase a target is significantly influenced by the sampling frequency \( F \). A higher \( F \) decreases the intervals between samples, which can accelerate data collection. However, this might also increase the total number of movements, consequently increasing the  moving time, as closely spaced samples may not provide comprehensive channel information.
These dynamics are crucial for the operational efficiency of the proposed chasing system, highlighting the need to assess the total tracking time across various \( F \) and \( \rho \) configurations.

The total tracking time, for a specified number of movements (\( M \)), sampling frequency (\( F \)), and tracking UAVs' velocity (\( v \)), is denoted as \( \tau(M,F) \) and calculated using the formula:
\begin{align}
\tau(M,F) = M \cdot \frac{L}{F} + \sum_{m=1}^{M} \frac{d_m}{v} \text{ [sec]},
\end{align}
where \( d_m \) denotes the distance traveled during the \( m \)-th movement, defined as \( d_m = |O_{m} - O_{m-1}|\). This formula breaks down the total time into channel observation time and the physical movement time between points.

\begin{figure}
     \centering
     \begin{subfigure}[b]{0.5\textwidth}
         \centering
         \includegraphics[width=8.5cm]{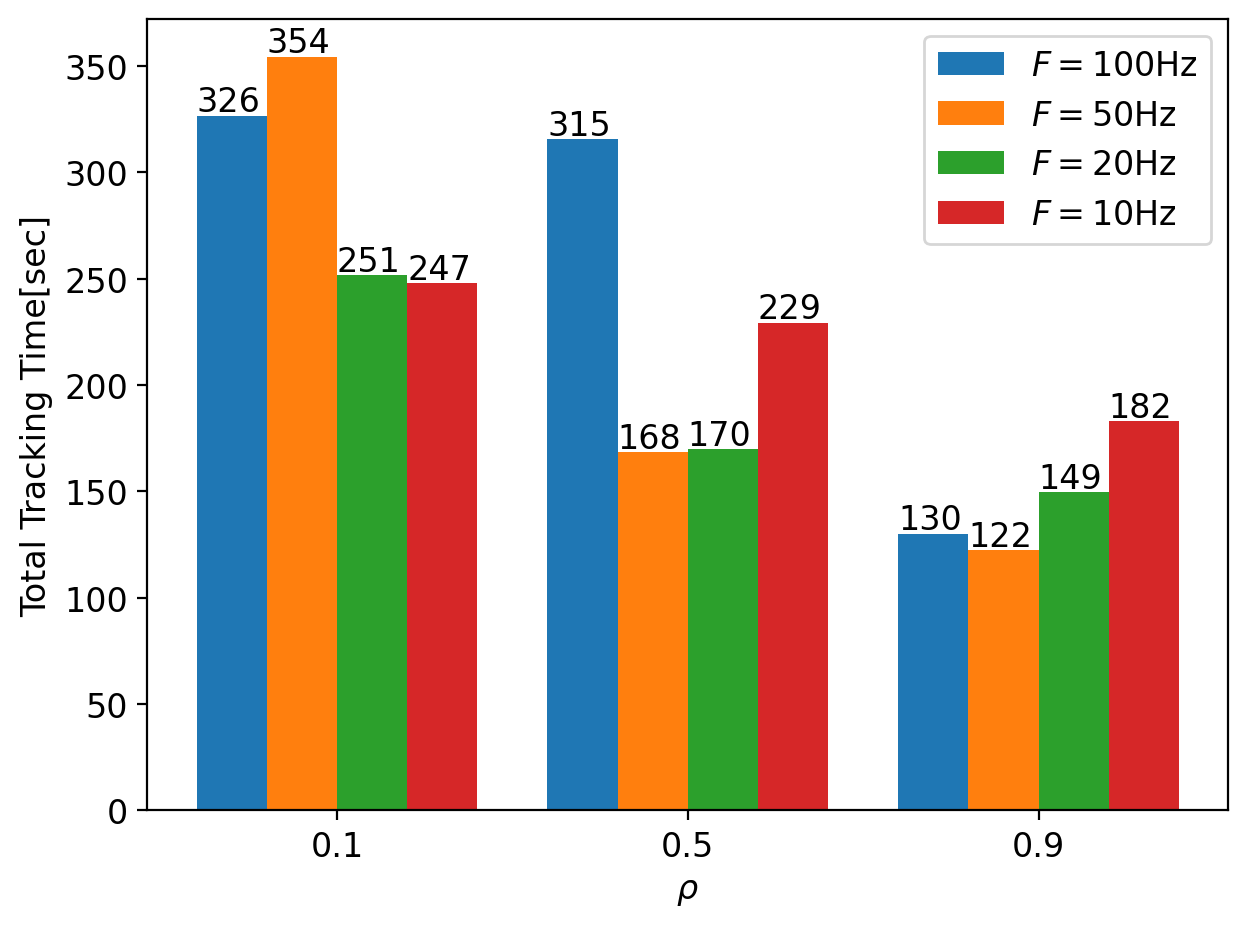}
         \caption{$v=2$m/s.}
     \end{subfigure}
     \begin{subfigure}[b]{0.5\textwidth}
         \centering
         \includegraphics[width=8.5cm]{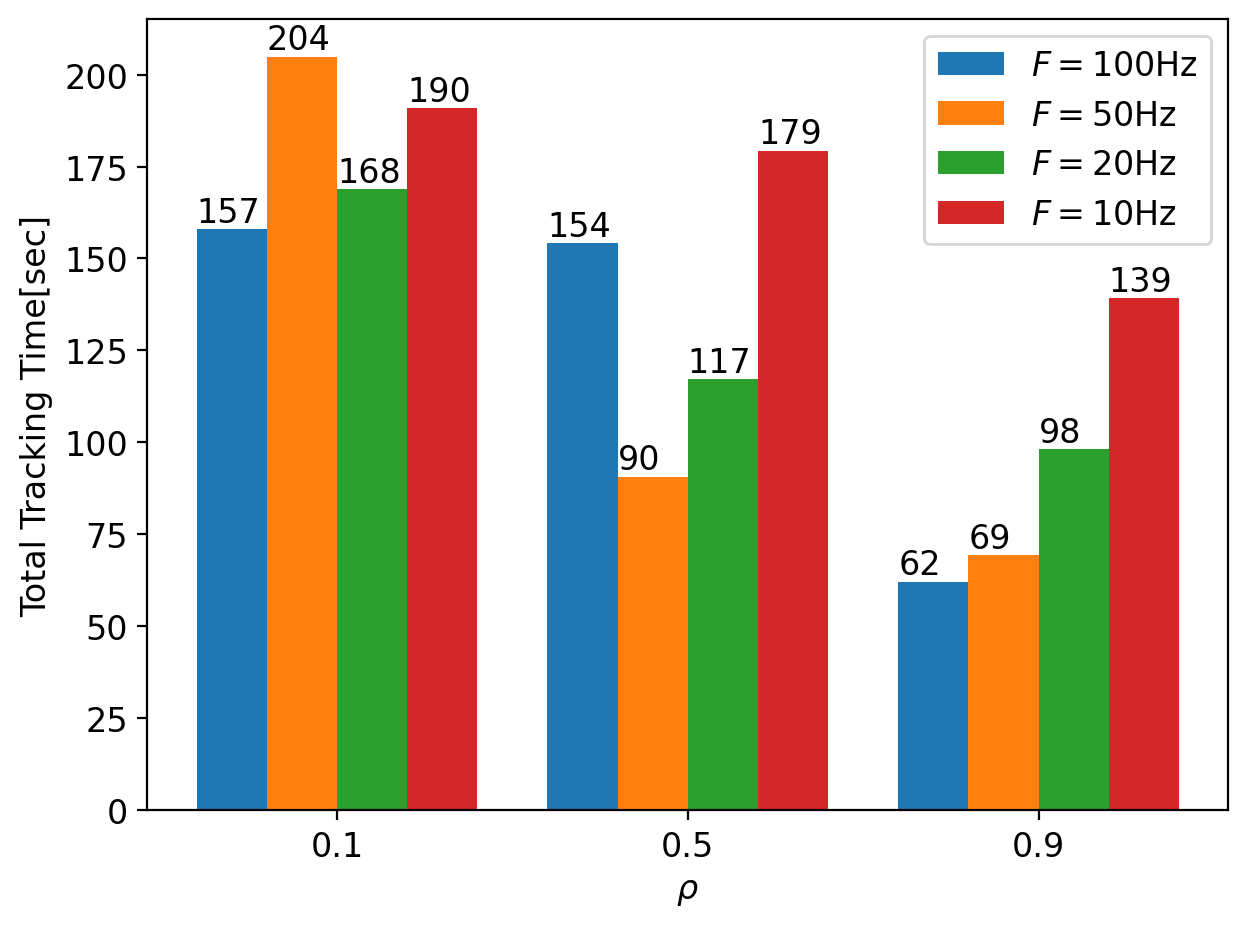}
         \caption{$v=5$m/s.}
     \end{subfigure}
        \caption{Average total tracking time.}
        \label{fig_total_tracking_time}
\end{figure}

Figs. \ref{fig_total_tracking_time} illustrate the average total tracking times for UAV velocities of \(v=2\) m/s and \(v=5\) m/s. The data clearly indicate that faster UAVs (\(v=5\) m/s) significantly reduce the total tracking time, primarily due to a decrease in the travel time component. Despite higher sampling frequencies (\(F\)) increasing the number of movements, as detailed in Fig. \ref{fig_step_percentile}, they can also result in shorter total tracking times due to reduced channel observation times, especially at higher correlation coefficients (\(\rho\)). For instance, at \(\rho=0.9\), increasing \(F\) from 10Hz to 100Hz reduces the total tracking time from 139 seconds to 62 seconds, even though the number of movements at the 90th percentile rises from 26 to 49, as illustrated in Fig. \ref{fig_step_percentile}.

It is evident that the optimal sampling frequency is dependent on both the velocity of the tracking UAVs and the spatial correlation coefficient. Faster UAV speeds and higher correlation coefficients can mitigate the drawbacks of increased movements caused by higher sampling frequencies, suggesting that adjustments to \(F\) should be made with careful consideration of both \(v\) and \(\rho\) to optimize tracking efficiency.

\section{Conclusions} \label{sec:5}
This paper proposed an innovative anti-UAV system designed to track and chase unauthorized UAVs using a sophisticated multidimensional swarm flight strategy, supported by the latest advancements in asynchronous deep reinforcement learning techniques. By enabling tracking UAV swarms to closely chase unauthorized target UAVs, this system significantly enhances the effectiveness of neutralizing such UAVs while concurrently minimizing the risk of collateral damage.

The core innovation of our system lies in the seamless integration of tracking and chasing functionalities by capturing communication signals between unauthorized UAVs and their controllers, traditionally handled as separate operations. A swarm flight strategy was also introduced to address the large-scale similarity issues of signals received by multiple antennas mounted on a single tracking UAV.

We have incorporated advanced channel modeling that accounts for spatial correlation and Doppler shifts, significantly enhancing the proposed system's responsiveness and adaptability to variable environmental dynamics. This modeling is crucial as it ensures the system maintains optimal performance under various operational conditions, as demonstrated by extensive simulations involving different channel sampling frequencies and spatial correlation coefficients.

Empirical results have consistently shown that the proposed system can achieve successful tracking and chasing within a defined criterion, irrespective of the sampling frequency and spatial correlation coefficients. This confirms the system's robust adaptability and reliability, establishing it as a valuable tool for both surveillance and active monitoring applications. The choice of channel sampling frequency ($F$) is crucial and should be carefully adjusted based on the tracking UAVs' velocity (\(v\)) and spatial correlation coefficient (\(\rho\)) to optimize the tracking efficiency of our system.

While our approach primarily relies on RSSIs, it can be extended to handle autonomous UAVs that do not emit RF signals by incorporating multi-sensor fusion techniques without altering the fundamental system structure. This enhancement will further enhance the system’s adaptability and expand its operational scope for broader UAV interception scenarios.

\bibliographystyle{IEEEtran}
\bibliography{ref}

\begin{thebibliography}{10}
\providecommand{\url}[1]{#1}
\csname url@samestyle\endcsname
\providecommand{\newblock}{\relax}
\providecommand{\bibinfo}[2]{#2}
\providecommand{\BIBentrySTDinterwordspacing}{\spaceskip=0pt\relax}
\providecommand{\BIBentryALTinterwordstretchfactor}{4}
\providecommand{\BIBentryALTinterwordspacing}{\spaceskip=\fontdimen2\font plus
\BIBentryALTinterwordstretchfactor\fontdimen3\font minus
  \fontdimen4\font\relax}
\providecommand{\BIBforeignlanguage}[2]{{%
\expandafter\ifx\csname l@#1\endcsname\relax
\typeout{** WARNING: IEEEtran.bst: No hyphenation pattern has been}%
\typeout{** loaded for the language `#1'. Using the pattern for}%
\typeout{** the default language instead.}%
\else
\language=\csname l@#1\endcsname
\fi
#2}}
\providecommand{\BIBdecl}{\relax}
\BIBdecl

\bibitem{droneapp2021}
``{IEEE Standard for Drone Applications Framework},'' \emph{IEEE Std
  1936.1-2021}, pp. 1--28, 2021.

\bibitem{Martinez2018}
O.~A. Martinez and M.~Cardona, ``{State of the Art and Future Trends on
  Unmanned Aerial Vehicle},'' in \emph{2018 International Conference on
  Research in Intelligent and Computing in Engineering (RICE)}, 2018, pp. 1--6.

\bibitem{Cisar2020}
P.~{\v C}isar, R.~Pinter, S.~M. {\v C}isar, and M.~Gligorijevi{\'c},
  ``{Principles of Anti-Drone Defense},'' in \emph{2020 11th IEEE International
  Conference on Cognitive Infocommunications (CogInfoCom)}, 2020, pp.
  000\,019--000\,026.

\bibitem{MarketsandMarkets}
\BIBentryALTinterwordspacing
``{UAV Market},'' Sep. 2022. [Online]. Available:
  \url{https://www.marketsandmarkets.com/Market-Reports/unmanned-aerial-vehicles-uav-market-662.html}
\BIBentrySTDinterwordspacing

\bibitem{Xu2020}
C.~Xu, X.~Liao, J.~Tan, H.~Ye, and H.~Lu, ``{Recent Research Progress of
  Unmanned Aerial Vehicle Regulation Policies and Technologies in Urban Low
  Altitude},'' \emph{IEEE Access}, vol.~8, pp. 74\,175--74\,194, 2020.

\bibitem{Tedeschi2024}
P.~Tedeschi, F.~A. Al~Nuaimi, A.~I. Awad, and E.~Natalizio, ``{Privacy-Aware
  Remote Identification for Unmanned Aerial Vehicles: Current Solutions,
  Potential Threats, and Future Directions},'' \emph{IEEE Transactions on
  Industrial Informatics}, vol.~20, no.~2, pp. 1069--1080, 2024.

\bibitem{FAA2023}
{Federal Aviation Administration}, ``Remote identification of drones,'' Federal
  Aviation Administration (FAA), 2023, available online:
  \url{https://www.faa.gov/uas/getting_started/remote_id}.

\bibitem{Sadovskis2022}
J.~Sadovskis and A.~Aboltins, ``Modern methods for {UAV} detection,
  classification, and tracking,'' in \emph{2022 IEEE 63th International
  Scientific Conference on Power and Electrical Engineering of Riga Technical
  University (RTUCON)}, 2022, pp. 1--7.

\bibitem{Zhang2023}
Y.~Zhang, C.~Wu, T.~Zhang, Y.~Liu, and Y.~Zheng, ``{Self-Attention Guidance and
  Multiscale Feature Fusion-Based UAV Image Object Detection},'' \emph{IEEE
  Geoscience and Remote Sensing Letters}, vol.~20, pp. 1--5, 2023.

\bibitem{You2023}
J.~You, Z.~Ye, J.~Gu, and J.~Pu, ``{UAV-Pose: A Dual Capture Network Algorithm
  for Low Altitude UAV Attitude Detection and Tracking},'' \emph{IEEE Access},
  vol.~11, pp. 129\,144--129\,155, 2023.

\bibitem{Lee2023}
H.~Lee, S.~Han, J.-I. Byeon, S.~Han, R.~Myung, J.~Joung, and J.~Choi,
  ``{CNN-Based UAV Detection and Classification Using Sensor Fusion},''
  \emph{IEEE Access}, vol.~11, pp. 68\,791--68\,808, 2023.

\bibitem{He2023}
M.~He, X.~Fang, D.~Huang, and Z.~Zhang, ``{A Hybrid Integration Method for
  Low-Observable Micro-UAV Trajectory Tracking by 2D MIMO Radar},'' in
  \emph{2023 38th Youth Academic Annual Conference of Chinese Association of
  Automation (YAC)}, 2023, pp. 809--813.

\bibitem{An2024}
Q.~An, C.~Yeh, Y.~Lu, Y.~He, and J.~Yang, ``{Time-Varying Angle Estimation of
  Multiple Unresolved Extended Targets via Monopulse Radar},'' \emph{IEEE
  Transactions on Aerospace and Electronic Systems}, pp. 1--19, 2024.

\bibitem{Jin2024}
W.-C. Jin, K.~Kim, and J.-W. Choi, ``{Adaptive Beam Control Considering
  Location Inaccuracy for Anti-UAV Systems},'' \emph{IEEE Transactions on
  Vehicular Technology}, vol.~73, no.~2, pp. 2320--2331, 2024.

\bibitem{KianiGaloogahi2017}
H.~Kiani~Galoogahi, A.~Fagg, and S.~Lucey, ``{Learning background-aware
  correlation filters for visual tracking},'' in \emph{Proceedings of the IEEE
  international conference on computer vision}, 2017, pp. 1135--1143.

\bibitem{Danelljan2016}
M.~Danelljan, A.~Robinson, F.~Shahbaz~Khan, and M.~Felsberg, ``{Beyond
  correlation filters: Learning continuous convolution operators for visual
  tracking},'' in \emph{Computer Vision--ECCV 2016: 14th European Conference,
  Amsterdam, The Netherlands, October 11-14, 2016, Proceedings, Part V
  14}.\hskip 1em plus 0.5em minus 0.4em\relax Springer, 2016, pp. 472--488.

\bibitem{Li2023a}
Y.~Li, H.~Zhang, Y.~Yang, H.~Liu, and D.~Yuan, ``{RISTrack: Learning Response
  Interference Suppression Correlation Filters for UAV Tracking},'' \emph{IEEE
  Geoscience and Remote Sensing Letters}, vol.~20, pp. 1--5, 2023.

\bibitem{Li2022}
J.~Li, D.~H. Ye, M.~Kolsch, J.~P. Wachs, and C.~A. Bouman, ``{Fast and Robust
  UAV to UAV Detection and Tracking From Video},'' \emph{IEEE Transactions on
  Emerging Topics in Computing}, vol.~10, no.~3, pp. 1519--1531, 2022.

\bibitem{Li2023}
J.~Li, W.~Zhang, Y.~Meng, S.~Li, L.~Ma, Z.~Liu, and H.~Zhu, ``{Secure and
  Efficient UAV Tracking in Space-Air-Ground Integrated Network},'' \emph{IEEE
  Transactions on Vehicular Technology}, vol.~72, no.~8, pp. 10\,682--10\,695,
  2023.

\bibitem{Park2021}
S.~Park, H.~T. Kim, S.~Lee, H.~Joo, and H.~Kim, ``{Survey on Anti-Drone
  Systems: Components, Designs, and Challenges},'' \emph{IEEE Access}, vol.~9,
  pp. 42\,635--42\,659, 2021.

\bibitem{Jurn2021}
Y.~N. Jurn, S.~A. Mahmood, and J.~A. Aldhaibani, ``{Anti-Drone System Based
  Different Technologies: Architecture, Threats and Challenges},'' in
  \emph{2021 11th IEEE International Conference on Control System, Computing
  and Engineering (ICCSCE)}, 2021, pp. 114--119.

\bibitem{3gpp_ts}
{3rd Generation Partnership Project (3GPP)}, ``{Evolved Universal Terrestrial
  Radio Access (E-UTRA); User Equipment (UE) Radio Transmission and
  Reception},'' Technical Specification 36.101 V17.13.0, 2024.

\bibitem{Brown2012}
T.~Brown, P.~Kyritsi, and E.~De~Carvalho, \emph{{Practical Guide to MIMO Radio
  Channel: with MATLAB Examples}}.\hskip 1em plus 0.5em minus 0.4em\relax
  Wiley, 2012.

\bibitem{matlab_spatial}
\BIBentryALTinterwordspacing
MathWorks, ``{Propagation Channel Models},'' May 2024. [Online]. Available:
  \url{https://kr.mathworks.com/help/lte/ug/propagation-channel-models.html}
\BIBentrySTDinterwordspacing

\bibitem{Rappaport2002}
T.~S. Rappaport, \emph{{Wireless Communications: Principles and Practice}s},
  2nd~ed.\hskip 1em plus 0.5em minus 0.4em\relax Prentice Hall, 2002.

\bibitem{matlab_temporal}
\BIBentryALTinterwordspacing
MathWorks, ``{Filter input signal through MIMO multipath fading channel},'' May
  2024. [Online]. Available:
  \url{https://kr.mathworks.com/help/comm/ref/comm.mimochannel-system-object.html#d126e189388}
\BIBentrySTDinterwordspacing

\bibitem{mniha16}
V.~Mnih, A.~P. Badia, M.~Mirza, A.~Graves, T.~Lillicrap, T.~Harley, D.~Silver,
  and K.~Kavukcuoglu, ``{Asynchronous Methods for Deep Reinforcement
  Learning},'' in \emph{Proceedings of The 33rd International Conference on
  Machine Learning}, 20--22 Jun 2016, pp. 1928--1937.

\end{thebibliography}

\newpage

\begin{IEEEbiography}[{\includegraphics[width=1in,height=1.25in,clip]{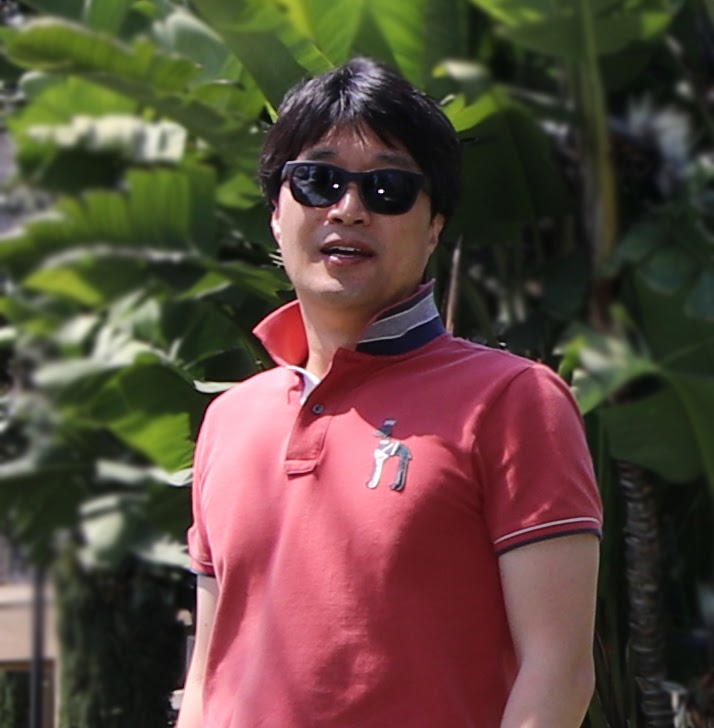}}]{Tae-Won Ban} (S'09-M'12) received the B.S. and M.S. degrees from the Department of Electronic Engineering, kyungpook National University, South Korea, in 1998 and 2000, respectively, and the Ph.D. degree from the Department of Electrical and Electronic Engineering, Korea Advanced Institute of Science and Technology (KAIST), Daejeon, South Korea, in 2010. He was a Researcher and a Network Engineer with Korea Telecom (KT), from 2000 to 2012. In KT, he researched 3G WCDMA, LTE, and Femto Systems. He was also responsible for traffic engineering and spectrum strategy. He is currently a Professor with the Department of AI and Information Engineering, Gyeongsang National University, South Korea. His current research interests include radio resource management for mobile communication systems and deep reinforcement learning.
\end{IEEEbiography}

\begin{IEEEbiography}[{\includegraphics[width=1in,height=1.25in,clip]{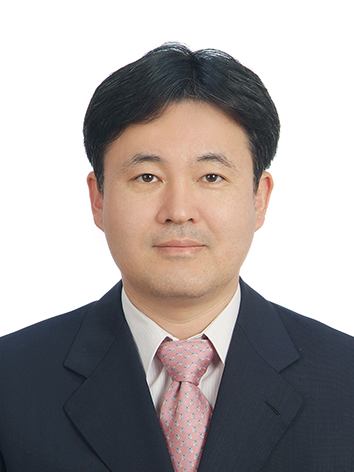}}]{Kyu-Min Kang} received the B.S., M.S., and Ph.D. degrees in electronic and electrical engineering from the Pohang University of Science and Technology (POSTECH), in 1997, 1999, and 2003, respectively. Since 2003, he has been with ETRI as a Principal Researcher. His current research interests include counter UAV technology, spectrum sharing, cognitive radio networks, digital signal processing, and high-speed digital transmission systems.
\end{IEEEbiography}

\begin{IEEEbiography}[{\includegraphics[width=1in,height=1.25in,clip]{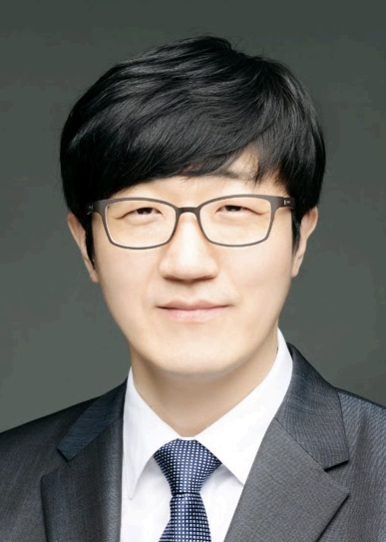}}]{Bang Chul Jung} (Senior Member, IEEE) received
the B.S. degree in electronics engineering from
Ajou University, Suwon, South Korea, in 2002,
and the M.S. and Ph.D. degrees in electrical and
computer engineering from Korea Advanced Institute for Science and Technology (KAIST), Daejeon,
South Korea, in 2004 and 2008, respectively. He
was a Senior Researcher/Research Professor with
the KAIST Institute for Information Technology
Convergence, Daejeon, from 2009 to 2010. He is
currently a Professor with the Department of Electronics Engineering, Chungnam National University, Daejeon. His research
interests include wireless communication systems, the Internet of Things (IoT)
communications, statistical signal processing, information theory, interference management, radio resource management, spectrum-sharing techniques,
and machine learning.,He received the Fifth IEEE Communication Society
Asia–Pacific Outstanding Young Researcher Award, in 2011, the Bronze Prize
in Intel Student Paper Contest, in 2005, the First Prize in the KAISTs Invention
Idea Contest, in 2008, and the Bronze Prize in Samsung Humantech Paper
Contest, in 2009. He was selected as a Winner of the Haedong Young Scholar
Award, in 2015, sponsored by the Haedong Foundation and given by KICS.
He has been selected as a Winner of the 29th Science and Technology Best
Paper Award, in 2019, sponsored by the Korean Federation of Science and
Technology Societies. He served as an Associate Editor for IEEE Vehicular
Technology Magazine, from 2020 to 2022, and is also a Senior Editor for
IEEE Vehicular Technology Magazine.
\end{IEEEbiography}

\end{document}